\documentclass[twocolumn]{aastex631}
\usepackage{physics}
\usepackage{mathrsfs}

\begin{document}

\title{High-Energy Neutrino Emission from a Radiatively Inefficient Accretion Flow \\\ with a three-dimensional GRMHD Simulation}

\author[0000-0001-8527-0496]{Tomohisa Kawashima}
\affiliation{National Institute of Technology, Ichinoseki College, Takanashi, Hagisho, Ichinoseki, Iwate, 021-8511, Japan}
\affiliation{Institute for Cosmic Ray Research, The University of Tokyo, 5-1-5 Kashiwanoha, Kashiwa, Chiba 277-8582, Japan}

\author[0000-0001-9064-160X]{Katsuaki Asano}
\affiliation{Institute for Cosmic Ray Research, The University of Tokyo, 5-1-5 Kashiwanoha, Kashiwa, Chiba 277-8582, Japan}



\begin{abstract}

The high-energy particle production in the accretion flow onto black holes can be a key physics to explain the high-energy neutrino background.
While the single-zone approximation has been commonly adopted in studies of the high-energy neutrino emission around black holes, the effects of the global plasma structure may be non-negligible. 
We carry out the first computations of cosmic-ray acceleration and high-energy neutrino emission via the hadronuclear ($pp$) interaction in global radiatively inefficient accretion flows and outflows around a supermassive black hole, using three-dimensional general relativistic magnetohydrodynamic simulation data. 
The Fokker-Planck equation for cosmic-ray protons is solved with
a phenomenological model for the energy diffusion coefficient  to express the turbulent acceleration in the sub-grid scale. The inhomogeneous and time variable
structure of the accretion flow
leads to a variety of particle energy distribution.
The spectral energy distributions (SEDs) of neutrinos emitted from the entire region are flatter than those calculated under the single-zone approximation. 
In our model, the neutrino emission originating from cosmic rays 
advected with the outflow rather than the inflow predominates the SEDs. Such galactic nuclei can be significant sources of cosmic rays in those galaxies.
\end{abstract}




\section{Introduction} \label{sec:intro}
High-energy neutrinos are anticipated to be a smoking gun for the acceleration of high-energy cosmic rays (CRs) \citep{Halzen_Hooper_2002review, Becker_2008review, Murase_Stecker_2023review}. 
In the last decade, IceCube has reported the detection and evidence of high-energy astrophysical neutrino background \citep{Arsten_etal_2013, IceCube_diffuse_2020, Halzen_Kheirandish_2022review}.

While 
the origin of
the diffuse astrophysical neutrino signal is still uncertain \citep[e.g.,][]{Murase_Waxman_2016},
the IceCube experiment data integrated over $\sim$ 10 years reveals the appearance of a neutrino hotspot in NGC 1068 with a significance of 4.2 $\sigma$ \citep{IceCube_NGC1068_2022}.
NGC 1068 is classified as a Seyfert II galaxy, which is one of the luminous classes of active galactic nuclei (AGNs) accompanying weak or no relativistic jets.
Recent analysis of X-ray bright Seyfert galaxies indicates that NGC 4151 and NGC 3079 might also be neutrino hotspots \citep{Neronov_etal_2024Seyfert, Abbasi_etal_2025_hardX_AGN, Abbasi_etal_2024arXiv_brightSeyfert}.
\cite{Murase_etal_2020} and  \cite{Murase_2022} demonstrate that CRPs, which are
injected via magnetic reconnections and accelerated via turbulence in compact coronae 
around
accretion disks, 
produce a soft neutrino SED due to energy loss via $p \gamma$ and Bethe-Heitler processes, 
while gamma-rays are hidden by $\gamma \gamma$ absorption.
The wind shock scenarios of the AGN neutrino emission have also been proposed \citep{InoueY_etal_2020, InoueS_etal_2022arXiv}.
However, the neutrino spectrum in NGC 1068 is significantly softer than that of the diffuse neutrino background. Another type of source may be required.

Various theoretical models on high-energy neutrino emissions from
AGNs have been proposed, to explain the diffuse astrophysical neutrino intensity \citep[e.g.,][]{Kimura_etal_2021}. 
In accretion flows onto supermassive black holes powering AGNs, turbulent acceleration of CR protons (CRPs) can occur \citep{Dermer_etal_1996}.
\cite{Kimura_etal_2015} apply the theory of turbulent acceleration of CRPs to radiatively inefficient accretion flows \citep[RIAFs:][]{Ichimaru_1977, Narayan_Yi_1994, Narayan_Yi_1995, Yuan_Narayan_2014review} in low luminosity AGNs (LLAGNs), and revealed that high-energy neutrino emissions via hadronuclear ($pp$) and photohadronic ($p \gamma$) interactions are expected;
in particular, the contribution of $pp$ interaction dominates the spectral energy distributions (SEDs) in the energy range $\lesssim 0.1 ~{\rm PeV}$.

In the above models of the high-energy neutrino emissions in AGNs, single-zone approximations (or the summation of the SEDs of different single zones)  are adopted.
Only a few studies using global simulations have been carried out for the neutrino emissions in the accretion flow.
For example, \cite{Kimura_etal_2019} study particle acceleration using Newtonian magnetohydrodynamic (MHD) simulation data, assuming the injection of very high-energy CRPs, whose Larmor radius is several times larger than the grid size in the simulations.
They find that the hard-sphere-like acceleration occurs through turbulence driven by magneto-rotational instability (MRI).
\cite{Rodriguez-Ramirez_etal_2019} compute the propagation of CRPs in the RIAF in Sgr A$^*$  using axisymmetric two-dimensional general relativistic MHD (GRMHD) simulation data
to show the SEDs of high-energy neutrinos and multi-wavelength photons including very-high-energy gamma-rays. 
In their study, however, the acceleration of CRPs is not solved, while the injection of CRPs is treated based on a semi-analytical estimate of  magnetic reconnection.
In present MHD simulations of turbulence in accretion flows, the grid size is much larger than the Larmor radius of sub-PeV particles. A phenomenological sub-grid model can be the first step in treating the particle acceleration process with global MHD simulations.

In this paper, we present 
the first attempt to compute CR acceleration and high-energy neutrino emission via the $pp$ interactions in global RIAFs and outflows around a supermassive black hole, by using three-dimensional GRMHD simulation data. 
Our purpose is observing the effects of inhomogeneity in the GRMHD simulations on the neutrino emission rather than determining the sub-grid model of the energy diffusion.
In \S \ref{sec:method}, we describe the computational method employed in our newly developed code $\nu$-\texttt{RAIKOU}, where a sub-grid model for particle acceleration is adopted.
In \S \ref{sec:result}, the results and analysis of the high-energy neutrino SEDs are presented.
We provide a discussion including a brief comparison between the diffuse neutrino intensity estimated from our global model and the one detected by the IceCube experiment in \ref{sec:discussion}.
Finally, \S \ref{sec:summary} is devoted to summarizing this work.

\section{Methods} \label{sec:method}
We calculate the spectra of high-energy neutrinos emitted via the $pp$ inelastic collision process in accretion flows and outflows by post-processing GRMHD simulation data.
CRPs are assumed to couple with the background fluid via microscopic wave-particle interaction.
We introduce tracer particles, 
which express the average position of a group of CRPs advected with the background flow. 
Their trajectories are computed along streamlines by updating the GRMHD simulation snapshot data over time. 
CRPs are assumed to be stochastically accelerated by kinetic-scale plasma turbulence (i.e., sub-grid scale turbulence in GRMHD simulations) in accretion flows and outflows. 
More details are provided in the following two subsections.

\subsection{GRMHD simulation data} \label{sec:GRMHD}
The time-dependent spatial distributions of MHD plasmas around a black hole are obtained by performing three-dimensional GRMHD simulations using the GR-radiation-MHD code \texttt{UWABAMI} \citep{Takahashi_etal_2016}.
In this paper, we use the GRMHD simulation data in \cite{RAIKOUcode}, where radiation effects are turned off for simplicity.

An overview of our GRMHD simulation is given below.
The dimensionless black hole spin parameter is set to be $a_* = 0.9375$.
The GRMHD equations are integrated in the modified Kerr-Schild coordinate system $(r, \theta, \varphi)$. 
The inner and outer outflowing boundaries of the simulation domain are imposed at $r_{\rm in} = 1.18 r_{\rm g}$ and $r_{\rm out} = 3 \times 10^3 r_{\rm g}$, respectively.
Here, $r_{\rm g} = GM_{\rm BH}/c^2$ is the gravitational radius, $G$ is the gravitational constant, $M_{\rm BH}$ is the black hole mass, and $c$ is the speed of light.
In the poloidal and azimuthal directions, the computational domain extends over $0 \le \theta \le \pi$ and $0 \le \varphi \le 2\pi$, respectively. 
The reflection and periodic boundary are imposed at $\theta = 0, \pi$ and $\varphi = 0, 2\pi$, respectively.
The simulation domain is divided into meshes of  $(N_r, N_{\theta}, N_{\varphi}) = (192, 120, 96)$. 
The resultant dimensionless magnetic flux  \citep[e.g.,][]{Tchekhovskoy_etal_2011}
threading the event horizon $\simeq 20$ in Gaussian-cgs units.
This state is sometimes referred to as "semi-MAD", which is an intermediate state between the SANE \citep[Standard and Normal Evolution, e.g., ][]{Narayan_etal_2012} and MAD \citep[Magnetically Arrested Disk, e.g.,][]{Tchekhovskoy_etal_2011}.

\begin{figure}
	\includegraphics[width=\columnwidth]{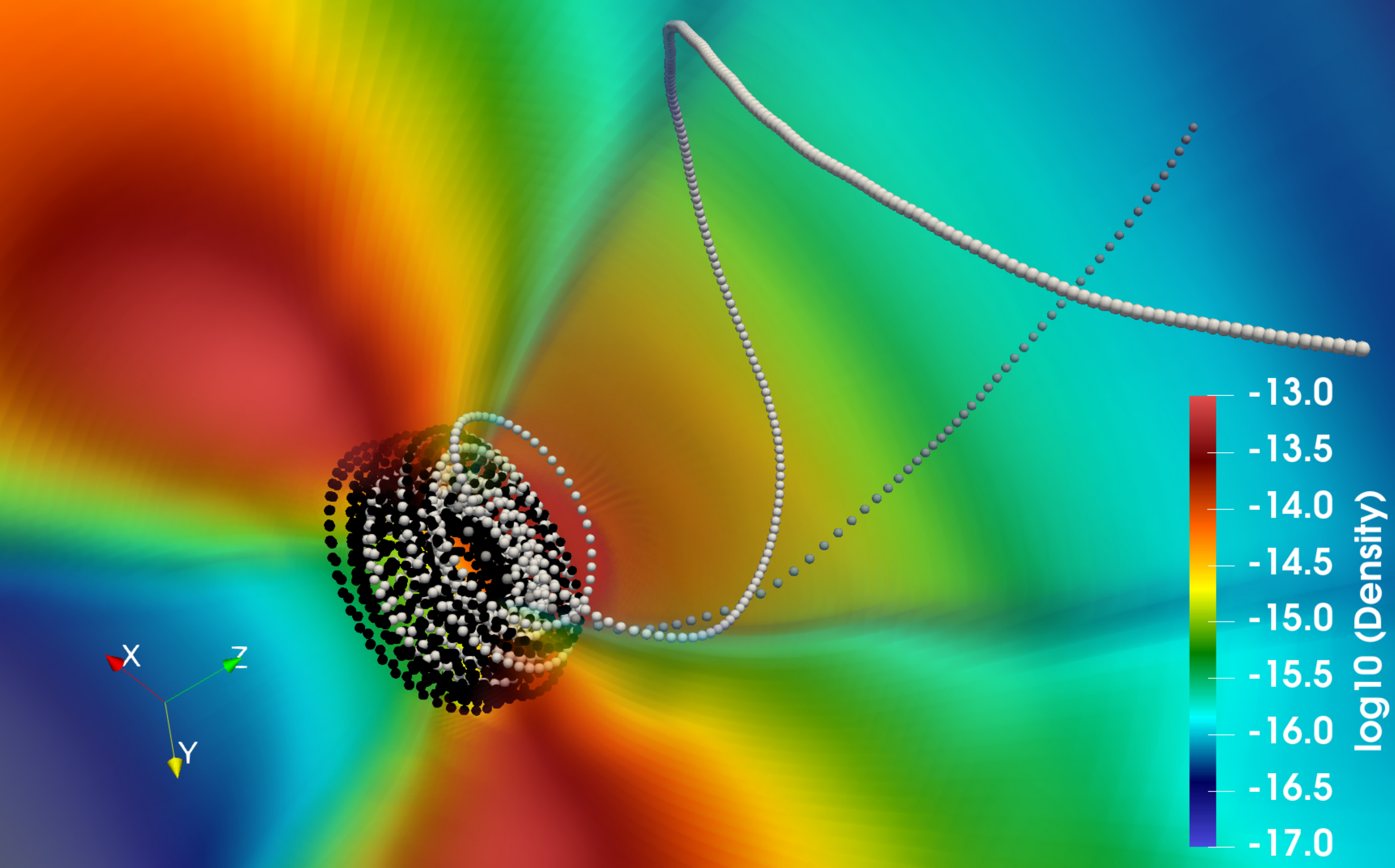}
    \caption{Volume rendering of the number density of 3D GRMHD simulation snapshot at $t=2\times 10^4 r_{\rm g}/c$. Three examples of the trajectories of CRPs are depicted by white, gray, and black dotted lines.}
    \label{fig:volume_rendering}
\end{figure}

Figure \ref{fig:volume_rendering} demonstrates a GRMHD simulation snapshot overlaid with three examples of CRP trajectories.
For the calculation of the CRP acceleration and neutrino emission, we process the simulation data as follows. We use the snapshot data from $ t = 8.5\times 10^{3} r_{\rm g}/c~ (\equiv t_{\rm start})$ to $5\times 10^4 r_{\rm g}/c~ (\equiv t_{\rm end})$, during which the mass accretion rate is quasi-steady. 
The GRMHD data coordinates are transferred from  modified Kerr-Schild to the Boyer-Lindquist coordinate system, so that inner boundary of the simulation domain is relocated to $\approx 1.36 r_{\rm g}$, which is just outside the outer event horizon ($\approx 1.35 r_{\rm g}$).
The outer boundary is also relocated to $\simeq 100 r_{\rm g}$, in order to remove the physically uncertain region outside the initial torus.
For the computations of the CRP evolution and neutrino emission, we neglect the strongly magnetized region $\sigma_{\rm mag} \equiv B'^2/(4 \pi \rho' c^2) > 1$  (the mass density $\rho'$ and magnetic field $B'$ are measured in the fluid rest frame), which is typically seen in the ``jet'' region, where a density floor is required.
Throughout this paper, the prime describes the physical quantities measured in the fluid rest frame.

Hereafter, we set the black hole mass and accretion rate to compute the neutrino spectra and to estimate the uncertainty neglecting the effects of the photohadronic proceeses, while the GRMHD simulations are scale free.
The black hole mass and time-averaged accretion rate are assumed to be $M_{\rm BH} = 10^8 M_{\odot}$ and ${\dot M}_{\rm in}c^2 = 10^{-2} L_{\rm Edd} \simeq 1.26 \times 10^{44} {\rm erg} ~ {\rm s}^{-1}$, respectively. Here, $L_{\rm Edd} (= 4 \pi GM_{\rm BH} m_{\rm p}/\sigma_{\rm T} c)$ is the Eddington luminosity, $m_{\rm p}$ is the proton mass, $\sigma_{\rm T}$ is the Thomson scattering cross section.
The mass outflow rate from the computation domain for the CRP acceleration and neutrino emission, which is described in \S  \ref{sec:method_CRP_neutrino}, is ${\dot M}_{\rm out} \simeq 3.41 \times 10^{44} {\rm erg} ~ {\rm s}^{-1}$, so that the mass outflow rate slightly dominates the mass accretion rate onto black hole in this simulation.

The black hole mass and accretion rate onto the black hole in our simulation is the same as one of the LLAGN models in \cite{Kimura_etal_2015}.
While considering a specific AGN is out of the scope of this paper,  we briefly mention that some of low-ionization nuclear emission line region (LINER), Seyfert galaxies with low Eddington ratio, radio galaxies, or BL Lac objects should have similar black hole masses and accretion rates. For example, NGC 1097, which is classified as a LINER, is estimated to have $M_{\rm BH} \simeq 1.2 \times 10^{8} M_{\odot}$ \citep{Lewis_Eracleous_2006_NGC1097_MBH} and ${\dot  M} \sim 6 \times 10^{-2} L_{\rm Edd}/c^2$ \citep[e.g.,][]{Nemmen_etal_2006_NGC1097mdot}. 
A BL Lac object Mrk 421 is also estimated to have a similar black hole mass $M_{\rm BH} \simeq 2-9 \times 10^{8} M_{\odot}$ \citep{Sahu_etal_2016}  and ${\dot  M} \sim 10^{-2} L_{\rm Edd}/c^2$ \citep[e.g.,][]{Celotti_etal_1998}\footnote{
We note that, while on-axis BL Lacs may not be the main sources of the diffuse neutrino, off-axis BL Lacs, which are not identified as blazars (e.g., radio galaxies), are still candidates for the neutrino source.}.
On the other hand, the neutrino luminosity claimed in TXS 0506+056 $\sim 10^{47} {\rm erg} ~ {\rm s}^{-1}$ requires super-Eddington accretion power \citep{Yang_etal_2025TXS0506}\footnote{While the the coronal emission is assumed in their model, there is a counter argument that it is difficult to reproduce the observed neutrino luminosity if the empirical coronal X-ray luminosity is adopted\citep{Fiorillo_etal_2025}. The origin and reliability of the neutrino emission from TXS 0506+056 is still under debate.}, which is out of the scope of this paper.

\subsection{CRP acceleration and neutrino emission} \label{sec:method_CRP_neutrino}
\subsubsection{Acceleration and injection of CRPs}
By post-processing the GRMHD snapshot data introduced in Section \ref{sec:GRMHD}, we compute trajectories and SEDs of CRPs using the newly developed code $\nu$-\texttt{RAIKOU}.
The details are explained below. 

We assume that CRP tracer particles move along the  streamlines of the GRMHD simulation snapshots by integrating the following equation: 
$\dd {r^i}/ \dd t = u^{i}/u^t$, where $u^t$ and $u^{i} (i=r, \theta, \varphi)$ are the time and spatial components of the fluid four-velocity in the Boyer-Lindquist coordinates, which are given by the GRMHD simulation data.
The trajectories of the CRPs are traced until they are either captured by the black hole or escape through the outer boundary of the computational domain.

We compute the distribution function per unit rest-mass of the background fluid
${\mathscr N}^{\prime}_{\rm CRP}(\epsilon_{\rm CRP}^{\prime}, t^{\prime})$ for each tracer particle, where $\epsilon^{\prime}_{\rm CRP}$ and $t^{\prime}$ are the energy of CRP and time, respectively.
The physical unit of ${\mathscr N}^{\prime}_{\rm CRP}$ is $[{\rm g}^{-1} ~ {\rm GeV}^{-1}]$. 
The Fokker--Planck equation of each CRP tracer particle in its fluid rest frame is given by
\begin{eqnarray}
\pdv{{\mathscr N}_{\rm CRP}^{\prime} (\epsilon_{\rm CRP}^{\prime},t^{\prime})}{t^{\prime}} =\pdv{}{\epsilon_{\rm CRP}^{\prime}} \left[D(\epsilon_{\rm CRP}^{\prime})\pdv{{\mathscr N}_{\rm CRP}^{\prime} (\epsilon_{\rm CRP}^{\prime},t^{\prime})}{\epsilon_{\rm CRP}^{\prime}}
\right] \nonumber \\
- \pdv{}{\epsilon_{\rm CRP}^{\prime}}
\left[\frac{2D(\epsilon_{\rm CRP}^{\prime})}{\epsilon_{\rm CRP}^{\prime}}
{\mathscr N}_{\rm CRP}^{\prime}(\epsilon_{\rm CRP}^{\prime},t^{\prime})
\right]  \nonumber \\ 
 + {\dot {\mathscr N}}_{\rm CRP}^{\prime ~ \rm comp}(\epsilon_{\rm CRP}^{\prime},t^{\prime})
 + {\dot {\mathscr N}}_{\rm CRP}^{\prime ~ \rm inj}(\epsilon_{\rm CRP}^{\prime},t^{\prime}), \label{eq:FP}
\end{eqnarray}
where $D(\epsilon_{\rm CRP}^{\prime})$, 
${\dot {\mathscr N}}_{\rm CRP}^{\prime ~ \rm comp}(\epsilon_{\rm CRP}^{\prime},t^{\prime})$
and ${\dot {\mathscr N}}_{\rm CRP}^{\prime ~ \rm inj}(\epsilon_{\rm CRP}^{\prime},t^{\prime})$ 
denote the energy diffusion coefficient, the  effect of adiabatic compression/expansion, and the injection rate of CRPs, respectively.

The energy diffusion of CRPs in RIAFs is highly uncertain. Here, we adopt a simple phenomenological model of the diffusion coefficient to demonstrate the variety of the CRP energy distribution. A detailed discussion of the justification of the diffusion coefficient is beyond the scope of this paper. For simplicity,
we adopt the hard-sphere approximation for the description of the diffusion coefficient \citep[e.g.][]{Teraki2019ApJ} given by
\begin{eqnarray}
D(\epsilon_{\rm CRP}^{\prime}) = K {\epsilon'}_{\rm CRP}^2.
\end{eqnarray}
The coefficient $K$ adjusts the acceleration timescale $t'_{\rm accel}=1/(2K)$. If $K$ is independent of CRPs, the CRP energy density increases exponentially. Such rapid evolution of the CRP energy density may not be realistic and hard to treat in numerical simulations. Thus, considering the back reaction of CRPs, we adopt a phenomenological model for $K$ to realize a relatively stable energy transfer from the turbulence to CRPs as follows.
As stronger turbulence may lead to a larger $K$, we assume that $t'_{\rm accel}$ is proportional to the velocity variation timescale as $t'_{\rm accel} \propto  |\vectorbold*{v}| / |{\rm D}\vectorbold*{v}/{\rm D}t^{\prime}|$, where ${\rm D}t'$ denotes the Lagrange derivative and $\vectorbold*{v}$ 
is the three-velocity of the MHD plasma measured in the Zero Angular Momentum Observer (ZAMO) frame. A part of the thermal energy in a grid should be the turbulence energy, which is numerically dissipated. So we can adopt a model with $K \propto U'_{\rm th}$, where $U'_{\rm th}$ is the thermal energy density. The Fokker--Planck equation implies $\partial U'_{\rm CRP}/ \partial t^{\prime} = 4 K U'_{\rm CRP}$, where $U'_{\rm CRP} \equiv \rho' \int d\epsilon_{\rm CRP}^{\prime} \epsilon_{\rm CRP}^{\prime} {\mathscr N}_{\rm CRP}^{\prime} (\epsilon_{\rm CRP}^{\prime})$. To have a form of $\partial U'_{\rm CRP}/ \partial t^{\prime} \propto U'_{\rm th}/t'_{\rm accel}$, we assume a phenomenological formula
\begin{eqnarray}
K = \frac{\eta_{\rm accel}}{4} \frac{|{\rm D}\vectorbold*{v}/{\rm D}t^{\prime}|}{|\vectorbold*{v}|} \frac{U'_{\rm th}}{U'_{\rm CRP}}, \label{eq:K_diff}
\end{eqnarray} 
which implies that the damping timescale of the turbulence energy density is relatively stable, i.e. independent of $U'_{\rm CRP}$.
The acceleration efficiency is adjusted by a free parameter $\eta_{\rm accel}$.
It should be noted that the acceleration timescale of CRPs in the early phase of the computation may be very short with this formalism. The resulting neutrino SEDs are, however, not expected to be critically affected by this initial CRP evolution.

We assume that CRPs are injected at $\epsilon_{\rm CRP}^{\prime} = 2m_{\rm p}c^2$  with the Gaussian distribution in energy space, with a dispersion of $\sigma_{\rm disp} = 0.1 m_{\rm p}c^2$.
So the normalized distribution function at injection is
\begin{eqnarray}
f_{\rm inj}(\epsilon_{\rm CRP}^{\prime}) 
&=& \frac{1}{\sqrt{2\pi \sigma_{\rm disp}^2}}\exp\left[-\frac{(\epsilon_{\rm CRP}^{\prime} - 2m_{\rm p}c^2)^2}{2\sigma_{\rm disp}^2}\right].
 \end{eqnarray}
Although the injection mechanism is unknown, magnetic reconnection can be a possible CRP injection mechanism. 
We assume that the injection rate is proportional to $\beta_{\rm mag}^{-2}$, 
where $\beta_{\rm mag}$ is the ratio of the thermal pressure to magnetic pressure, 
since the injection rate via magnetic reconnection may be proportional to the magnetic energy density and the Alfv$\acute{e}$n crossing time of the current sheets, whose thickness is proportional to the Larmor radius for the case of the plasma with a guide field \citep{Horiuchi_Sato_1997}. 
We impose an upper limit of the $\beta_{\rm mag}^{-2}$ to be $\max(1, \beta_{\rm mag}^{-2})$ to relax unexpectedly prompt injections in the highly magnetized region.   
Similarly to the diffusion coefficient, we assume that the reconnection rate is proportional to $|{\rm D}\vectorbold*{B}'/{\rm D}t^{\prime}|/|\vectorbold*{B}'| $. Then, the injection term can be written as
\begin{eqnarray}
{\dot {\mathscr N}}_{\rm CRP}^{\prime ~\rm inj}(\epsilon_{\rm CRP}^{\prime},t^{\prime}) 
&=& \frac{\eta_{\rm inj}}{m_{\rm p} }
\max(1, \beta_{\rm mag}^{-2})\frac{|{\rm D}\vectorbold*{B}'/{\rm D}t^{\prime}|}{|\vectorbold*{B}'|} f_{\rm inj}(\epsilon_{\rm CRP}^{\prime}).
 \end{eqnarray}  
The constant $\eta_{\rm inj}$ is introduced as a free parameter regulating the CRP injection rate. 

The energy diffusion and particle injection terms in Equation (\ref{eq:FP}) are solved using the Green's function \citep{Becker_etal_2006, Asano_Meszaros_2016}.
The adiabatic compression/expansion term is handled  by shifting the energy of the distribution function according to ${\rm D} \epsilon_{\rm CRP}^{\prime} / {\rm D}t^{\prime} = - (1/3) \epsilon_{\rm CRP}^{\prime} ({\vectorbold* \nabla} \cdot {\vectorbold* v})$, which corresponds to the energy shift of $ \Delta {\epsilon}_{\rm CRP}^{\prime}/{\epsilon}_{\rm CRP}^{\prime}=(1/3) \Delta \rho'/\rho'$.
Using $U'_{\rm th}$, 
$\vectorbold*{v}$,
$\Delta \vectorbold*{v}$, 
$\vectorbold*{B}'$,
$\Delta \vectorbold*{B}'$, $\beta_{\rm mag}$, and $\Delta \rho'$ in the GRMHD simulation data at the mesh where the tracer particle resides, 
we follow the evolution of the CRP energy distribution.
We note that the injection, compression, and the injection terms are solved with the ceiling value of $U_{\rm CRP} \le 0.2 ~U_{\rm th}$, since the CRP feedback effects on the GRMHD dynamics are not implemented in our simulations.
This upper limit value for the CRP energy density is the same as that adopted to the single zone models in \cite{Kimura_etal_2015}.

We input the total number $n_{\rm tot} = 11520$ of CRP tracer particles at $r\approx 10 r_{\rm g}$ and $t = 8.5 \times 10^3 r_{\rm g}/c$, for simplicity.
The time evolution of ${\mathscr N}_{\rm CRP}^{\prime}$ of each tracer particle is solved within the energy range $1 \le \epsilon_{\rm CRP}^{\prime}/m_{p}c^2  \le 10^{10}$ (i.e.,  $9.38 \times 10^{8} {\rm eV} \le \epsilon_{\rm CRP}^{\prime} \le 9.38 \times 10^{18} {\rm eV}$), which is divided into $5600$ energy bins. 
In our fiducial run, we set the acceleration and injection efficiency parameters to be $\eta_{\rm accel} = 3 \times 10^{-5}$ and $\eta_{\rm inj} =  10^{-3}$.
For the parameter studies, we also examine the cases with $\eta_{\rm accel} = 10^{-5}$ and $10^{-4}$ as well as $\eta_{\rm inj} = 3 \times 10^{-4}$ and $3 \times 10^{-3}$.

\subsubsection{Emission of high-energy neutrinos}
The neutrino production rate via $pp$ interactions is calculated with the same method in \citet{Nishiwaki_etal_2021}. We use the total cross-section of pion production in \citet{Kamae_etal_2006inclusive_crosssection, Kamae_etal_2007inclusive_crosssection}, and the neutrino energy distribution per pion in \citet{Kelner_2006}.
At each moment, the neutrino spectral emissivity ${\mathscr E}'_{\nu} (\epsilon'_{\nu})$ per unit mass  are transformed from fluid rest frame to the observer frame by taking into account the gravitational redshift. For simplicity, the effect of the special relativistic Doppler effects is neglected in this study. 
We integrate the neutrino SEDs (per unit mass) over the observer time.
Finally, the time-averaged SEDs are computed by normalizing the weight of the CRPs as follows:
\begin{eqnarray}
    \epsilon_{\nu} L_{\epsilon_{\nu}}^{\rm (inflow)} &=& {\dot M}_{\rm in} \frac{\sum_{n_{\rm in}} \int \dd t ~ \epsilon_{\nu} {\mathscr E}_{\nu}^{(n_{\rm in})} \left(\epsilon_{\nu}\right) w^{(n_{\rm in})}}{\sum_{n_{\rm in}}w^{(n_{\rm in})}},     \label{eq:in} \\
    \epsilon_{\nu} L_{\epsilon_{\nu}}^{\rm (outflow)} &=& {\dot M}_{\rm out} \frac{\sum_{n_{\rm out}} \int \dd t ~ \epsilon_{\nu} {\mathscr E}^{n_{\rm out}}_{\nu} \left(\epsilon_{\nu}\right) w^{(n_{\rm out})}}{\sum_{n_{\rm out}} w^{(n_{\rm out})}},
    \label{eq:out}
\end{eqnarray}
where the superscripts $(n_{\rm in})$ and $(n_{\rm out})$ denote the identification number of the tracer particles that eventually accrete onto the black hole and escape from the calculation box, respectively. 
The weight $w^{(n_{\rm in})}$ and $w^{(n_{\rm out})}$ are proportional to the total mass of the thermal proton in the initial mesh.
We adopt the time-averaged value of the mass accretion rate ${\dot M}_{\rm in}$ and outflow rate ${\dot M}_{\rm out}$ in our GRMHD simulation data.

\begin{figure}[t]
	\includegraphics[width=\columnwidth]{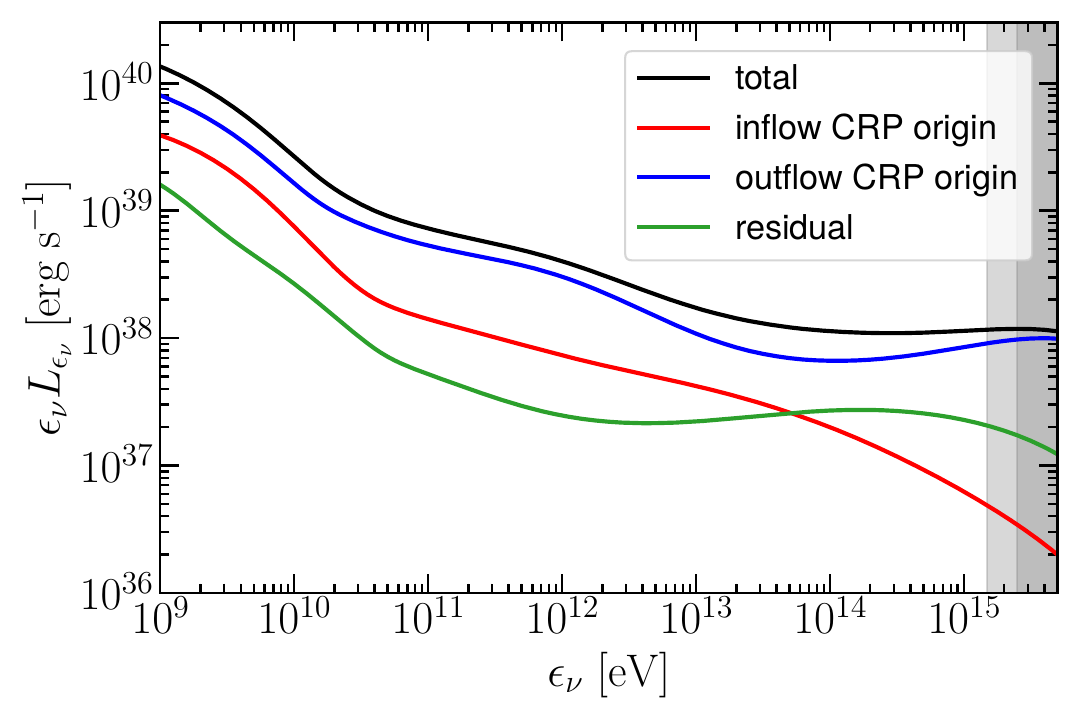}    
    \caption{The time-averaged SED of the all-flavored high-energy neutrinos. The total SED is also decomposed into contributions from different fates of CRPs: inflow (i.e., black-hole-captured) CRPs, outflow (i.e., escape) CRPs, and residual CRPs. 
    The light gray region reflects the uncertainty due to the parent CRPs whose Larmor radius is larger than the grid size, and the dark gray region
    corresponds to the uncertainty due to neglecting the photohadronic interactions
    in addition to the issue of the Larmor radius. The mass accretion rate is set to be ${\dot M}_{\rm in} = 10^{-2} L_{\rm Edd}/c^2$.} 
    \label{fig:Neutrino_SED_ave_sum_flux_woObs}
\end{figure}

\begin{figure*}
 	\includegraphics[width=\textwidth]{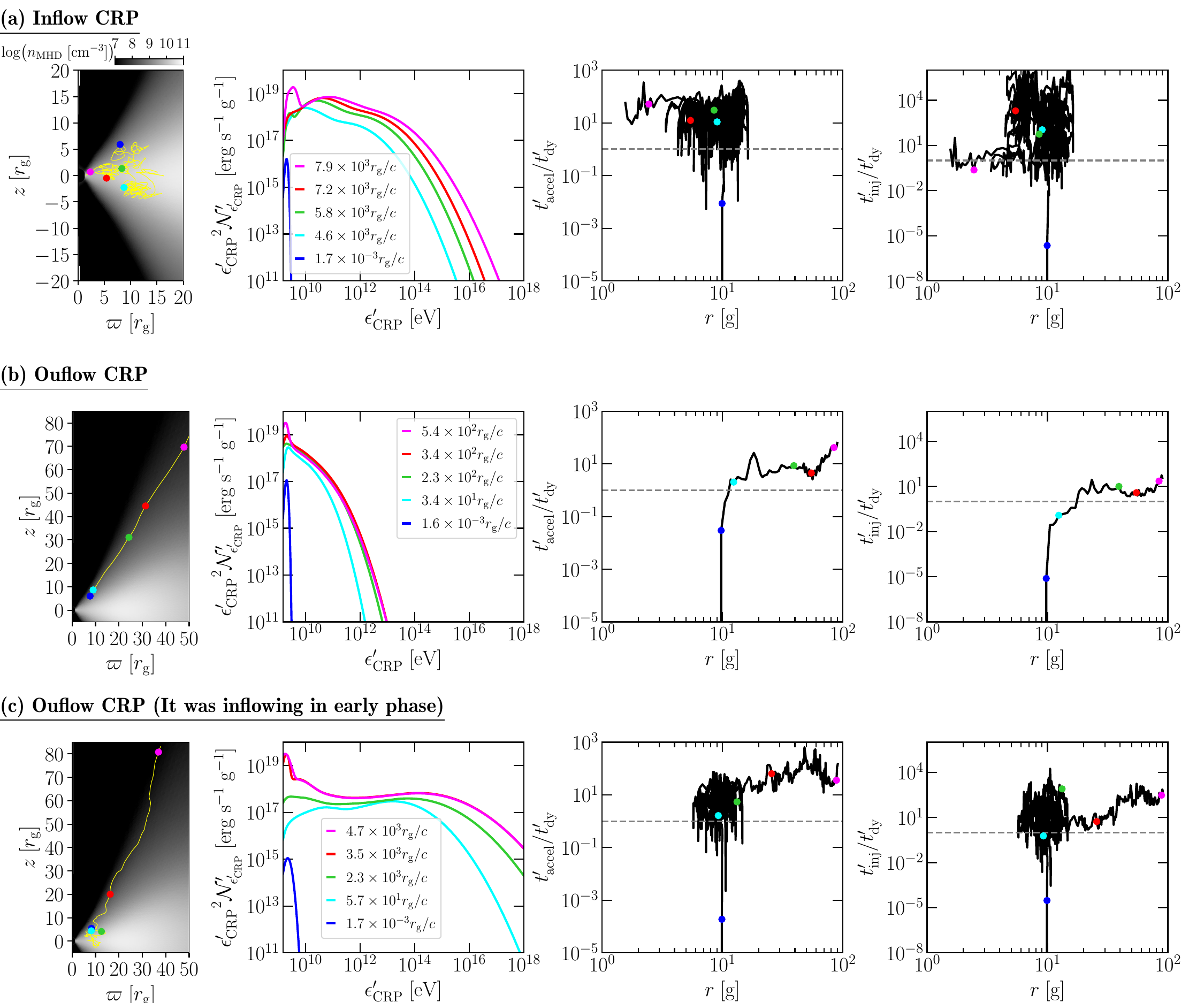}
    \caption{Three examples of time sequences of CRP energy distributions, and timescales 
    along tracer particles. 
    The top, middle, and bottom rows show the results for (a) inflow CRPs, (b) outflow CRPs, and (c) outflow CRPs that undergo accretion flows in the early phase of the computation, respectively. First column: trajectories of the CRP tracer particles overlaid onto the spatial distribution of time- and azimuthally-averaged proton number density on the poloidal plane in the GRMHD simulation. Second column: the CRP energy distribution functions. The third column: time evolution of the acceleration timescale $t_{\rm accel}^{\prime}$ normalized by the dynamical timescale $t_{\rm dy}^{\prime}$ measured in the fluid rest frame, as a function of the temporal location radius $r$. The fourth column: the same as the third column but for
    the injection timescale $t_{\rm inj}^{\prime}$. 
    The filled circles with the different colors
    represent 
    the time $t - t_{\rm start}$, 
    which are denoted in the second column. }
    \label{fig:CRP_trajectory_SED}
\end{figure*}

\section{Results}  \label{sec:result}
\subsection{Overview of the resulting neutrino SEDs} \label{sec:result_overview}
In the single-zone approximation, the hard-sphere type acceleration leads to a hard particle spectrum; especially in the steady state we have ${\mathscr N}_{\rm CRP} \propto \epsilon_{\rm CRP}^{-1}$ \citep[see, e.g.,][]{2014ApJ...780...64A,Kimura_etal_2015, Asano_Meszaros_2016}.
On the other hand, when the acceleration timescale is longer than the elapsed time, the particle spectrum becomes very soft, showing an exponential cutoff just above the spectral peak energy near the injection energy.
However, the time-averaged all-flavored neutrino spectra in Figure \ref{fig:Neutrino_SED_ave_sum_flux_woObs} extend to energies above PeV, and are softer than $\epsilon_{\nu}^{-2}$.
In addition, our results show that the spectrum gradually becomes harder with energy.

We also decompose the neutrino SED by classifying the fates of the parent CRPs: CRPs inflowed and outflowed as described in eqs. (\ref{eq:in}) and (\ref{eq:out}). 
It is found that the neutrinos generated by outflow CRPs are slightly more dominant than those produced by inflow CRPs.
This is a non-trivial discovery found from the 3D GRMHD simulation. If the accretion flow models for the neutrino sources are valid, the  outflow (i.e., the disk wind in this study) should be an important cosmic ray supplier.
We also note that, since our computations are terminated at $5\times 10^4 r_{\rm g}/c$, some CRPs remain within the simulation domain without escaping or accreting.
In Figure \ref{fig:Neutrino_SED_ave_sum_flux_woObs}, the neutrino flux from such residual CRPs is calculated averaged over the computation timescale, $4.15\times 10^4 r_{\rm g}/c$. As the neutrino emission from residual CRPs may last further changing its flux, this component brings uncertainty to the total flux.
However, the contribution of neutrino emission from these residual CRPs to the total SEDs is minor at the end of the simulation, and we expect that the residual CRPs do not significantly affect our conclusions.

\begin{figure*}[ht!]
	\includegraphics[width=\textwidth]{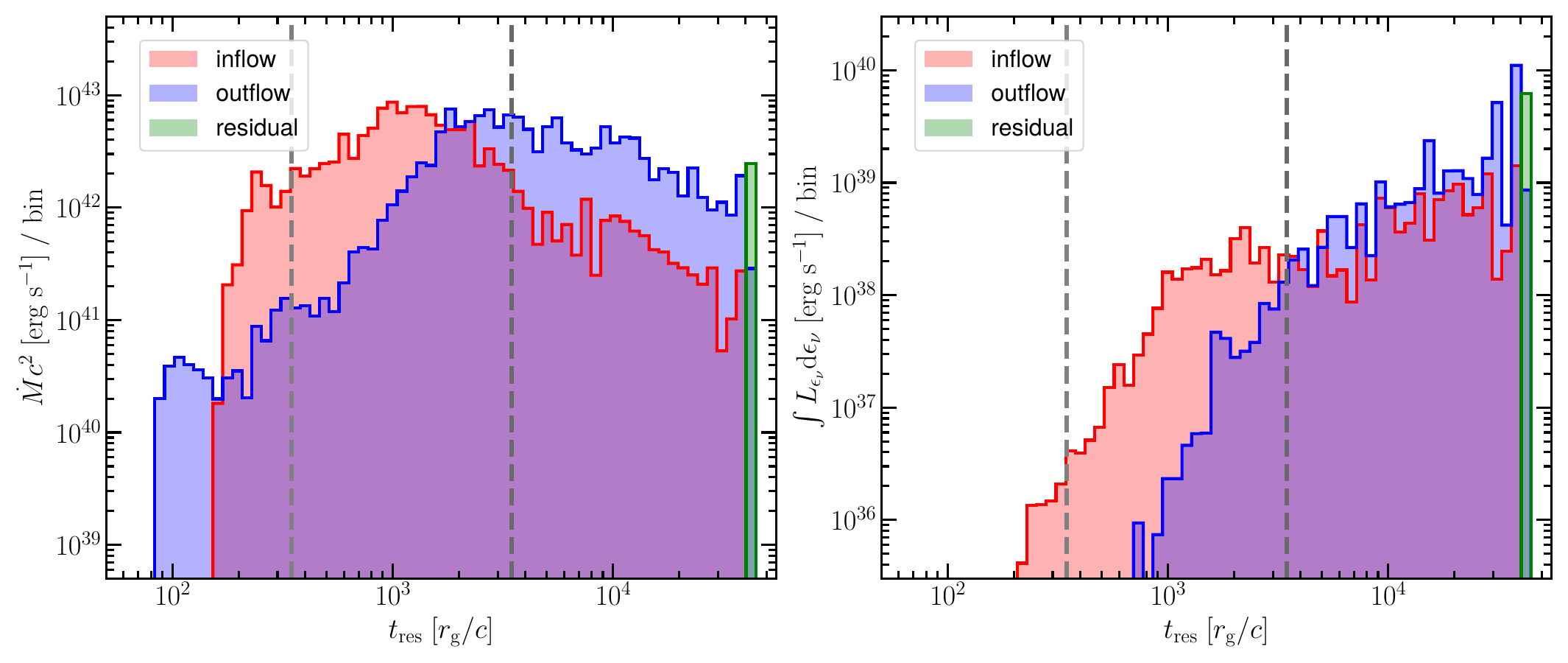}    \caption{Distributions of ${\dot M} c^2$ (left) and the neutrino luminosity $\int L_{\epsilon_{\nu}} \dd \epsilon_{\nu}$ as a function of the residence time in the simulation domain $t_{\rm res} \equiv t_{\rm esc} - t_{\rm start} $.
    We note that the residual component represents the duration of $t_{\rm end} - t_{\rm start}$ integrated in the fluid rest frame, for convenience. The vertical dashed lines indicate the viscous accretion timescale for $\alpha = 0.01$ (light gray) and 0.1 (dark gray).}    \label{fig:histogram_inflow_outflow}
\end{figure*}

Here, we note two caveats regarding our resultant neutrino SEDs. First, our assumption that CRPs  advected with the background fluid is valid only for CRPs whose Larmor radius is smaller than the mesh size. 
This condition is violated for CRPs with energies  $\gtrsim 3 \times 10^{16} ~ {\rm eV}$.
Second, the effects of photohadronic interactions are not incorporated in this study. 
The estimated CRP energy range in which $p\gamma$ interactions dominates over  $pp$ collision processes is $\gtrsim 5 \times 10^{16} ~ {\rm eV}$ (see, Appendix \ref{sec:app1}).
We assume that the X-ray luminosity is $\sim 10^{40} {\rm erg} ~ {\rm s}^{-1}$ \citep{Kimura_etal_2015} to estimate the timescale of the $p\gamma$ interactions.
These two uncertainties are shown by the gray shaded regions in Figure \ref{fig:Neutrino_SED_ave_sum_flux_woObs}.

\subsection{Analysis of inflow and outflow CRPs}
The neutrino SED consists of a superposition of different SED profiles. Figure \ref{fig:CRP_trajectory_SED} shows three examples of the time sequences of CRP energy distributions, acceleration timescales, and injection timescales  evolved along the trajectory of the tracer particles.
We select three trajectories, i.e., (a) an inflow CRP, (b) an outflow CRP, and (c) an outflow CRP that was effectively accelerated inside the accretion flow in the early stage of the simulation.
From the figures, the variety of the energy distributions is confirmed. The superposition of neutrino emission from those different CRP groups results in the soft neutrino spectrum.

We discuss the spectral evolution of CRPs in more detail, along with the timescales of acceleration and injection depicted in Figure \ref{fig:CRP_trajectory_SED}.
The inflowing trajectory
exhibits a turbulent-like signature inside the accretion flow in the leftmost panel in the top row.
The CRPs are efficiently accelerated over time by our phenomenological kinetic-scale turbulence.
The acceleration timescale frequently becomes
shorter than the dynamical timescale, especially in the early stage of the computation, as shown in the second panel from the right.
The high efficiency in the early stage is due to the initial low value of $U'_{\rm CR}$ ($D \propto {U'}_{\rm CR}^{-1}$).
The acceleration and dynamical timescales are estimated as $t_{\rm accel}^{\prime} =  1/(2K)$ 
and $t_{\rm dy}^{\prime} = r\gamma/|v^r|$ at each point along the CRP trajectory, where $\gamma$ is the Lorentz factor of the fluid and $v^r$ is the radial fluid velocity in the ZAMO frame.
We use  $t_{\rm dy}^{\prime}$ to roughly compare this with the acceleration and injection timescales.
In practice, the turbulent-like MHD plasma motion appears so that the dynamical timescale will be longer than this simple estimate. 
For another argument based on time-averaged quantities, we also estimate a time-averaged acceleration timescale $\langle t_{\rm accel}^{\prime} \rangle \equiv \exp\left[\int_{\rm t_{\rm start}^{\prime}}^{t_{\rm esc}^{\prime}} \log(t_{\rm accel}^{\prime})dt^{\prime} /\left( t_{\rm esc}^{\prime} - t_{\rm start}^{\prime} \right)\right]$, where $t_{\rm start}^{\prime}$ and $t_{\rm esc}^{\prime}$ are the time starting to trace CRPs and that of CRPs escaping the simulation domain measured in the fluid rest frame, respectively. It is found that $\langle t_{\rm accel}^{\prime} \rangle / \left( t_{\rm esc}^{\prime} - t_{\rm start}^{\prime} \right) \simeq $ 1.20,  4.58, and 1.18 for the (a) inflow CRP, (b) outflow CRP, and (c) outflow (initially inflow) CRP in Figure \ref{fig:CRP_trajectory_SED}, respectively. 
These estimates are consistent with the fact that the spectra of (a) and (c) are harder than that of (b).

Additionally, the adiabatic compression contributes to the energy gain of CRPs in the later stage of the computation ($t - t_{\rm start} ~ \gtrsim ~ 5\times 10^3 r_{\rm g}/c)$, as CRPs accrete towards the black hole.

With respect to the CRP injection, the injection timescale is evaluated by  $t_{\rm inj}^{\prime} = \int {\mathscr N}_{\rm CRP}^{\prime} d \epsilon'_{\rm CRP} / {\dot {\mathscr N}}_{\rm CRP}^{\prime  ~{\rm inj}}$.
The CRP injection occurs mainly in the transition layer of accretion flows and outflows, where $\beta_{\rm mag}$ is lower. 
This is the reason why the injection efficiently  in Fig. \ref{fig:CRP_trajectory_SED} (a) takes place in the early stage ($t - t_{\rm start} \sim 1.7\times 10^{-3} r_{\rm g}/c)$ and the late stage ($t - t_{\rm start} \sim 7.9\times 10^{3} r_{\rm g}/c)$ of the computations, during which the CRPs move along the streamlines near the transition layers.

In the middle panels of Figure \ref{fig:CRP_trajectory_SED}, the characteristics of outflow CRPs are presented. 
In this example, as the trajectory is relatively smooth (note that we plot only the poloidal trajectory), the acceleration timescale is longer than the dynamical timescale except in the early phase of the computation. 
Thus, the particle acceleration is less efficient.
The CRP distribution function just before escaping the simulation domain ($t-t_{\rm start} = 5.4 \times 10^2 r_{\rm g}/c)$ is slightly softer than that early ($t-t_{\rm strart} = 3.4 \times 10^2 r_{\rm g}/c)$, due to the adiabatic-expansion energy loss in the outflow region.

On the other hand, a significant fraction of outflow CRPs also undergo the turbulent accretion flows in the early phase of the computations, as shown in the bottom panels of Figure \ref{fig:CRP_trajectory_SED}.
The CRPs are efficiently accelerated inside the accretion flow until $t -t_{\rm start} \lesssim 3 \times 10^3 r_{\rm g}/c$.
This type of outflow CRPs also significantly contributes to the neutrino emission, as the target thermal proton density is high in the accretion flow.  
After this phase, the CRPs are shifted to the outflow.

While the steady state solution for the hard-sphere acceleration produces a CRP spectrum harder than $\epsilon_{\rm CRP}^{-2}$ \citep[e.g.][]{2014ApJ...780...64A}, such hard spectra are not seen in Figure \ref{fig:CRP_trajectory_SED}. The non-steady evolutions of the CRP injection rate and energy diffusion lead to a softer spectrum. The variety of such soft CRP spectra synthesizes the average neutrino spectrum in Figure \ref{fig:Neutrino_SED_ave_sum_flux_woObs}.

Figure \ref{fig:histogram_inflow_outflow} presents statistical analyses for the residence time.
The typical 
viscous-accretion timescale can be written as $t_{\rm vis} \sim {r^2/\nu_{\rm vis}}$, where $\nu_{\rm vis}= \alpha c_{\rm s} H$ is the kinetic viscosity assuming $\alpha$ prescription \citep{Shakura_Sunyaev_1973, Lynden-Bell_Pringle_1974} with the sound speed in the relativistic fluid $c_{\rm s} \sim c/\sqrt{3}$ and the scale height of RIAFs $H \sim r/2$  \citep[e.g.,][]{Manmoto_etal_1997}.
Adopting typical $\alpha$ values of $\alpha= 0.01$ and $0.1$, which are suggested in MHD simulations \citep[e.g.,][]{Machida_Matsumoto_2003,Hawley_etal_2011},
one can find that most of the inflow CRPs are distributed in this typical residence timescale.
The right panel of Figure \ref{fig:histogram_inflow_outflow} shows that CRPs  with a longer residence time contribute more significantly to neutrino emission.
A large fraction of outflow CRPs remain for a longer timescale $(t_{\rm esc}- t_{\rm start} \gtrsim 10^3 r_{\rm g}/c)$ than the typical residence time for inflow CRPs, which is consistent with the dominance of the neutrino emission originating from the outflow CRPs in the SED as shown in Figure \ref{fig:Neutrino_SED_ave_sum_flux_woObs}.

\subsection{Dependence of acceleration and injection efficiencies of CRPs}

\begin{figure}
	\includegraphics[width=\columnwidth]{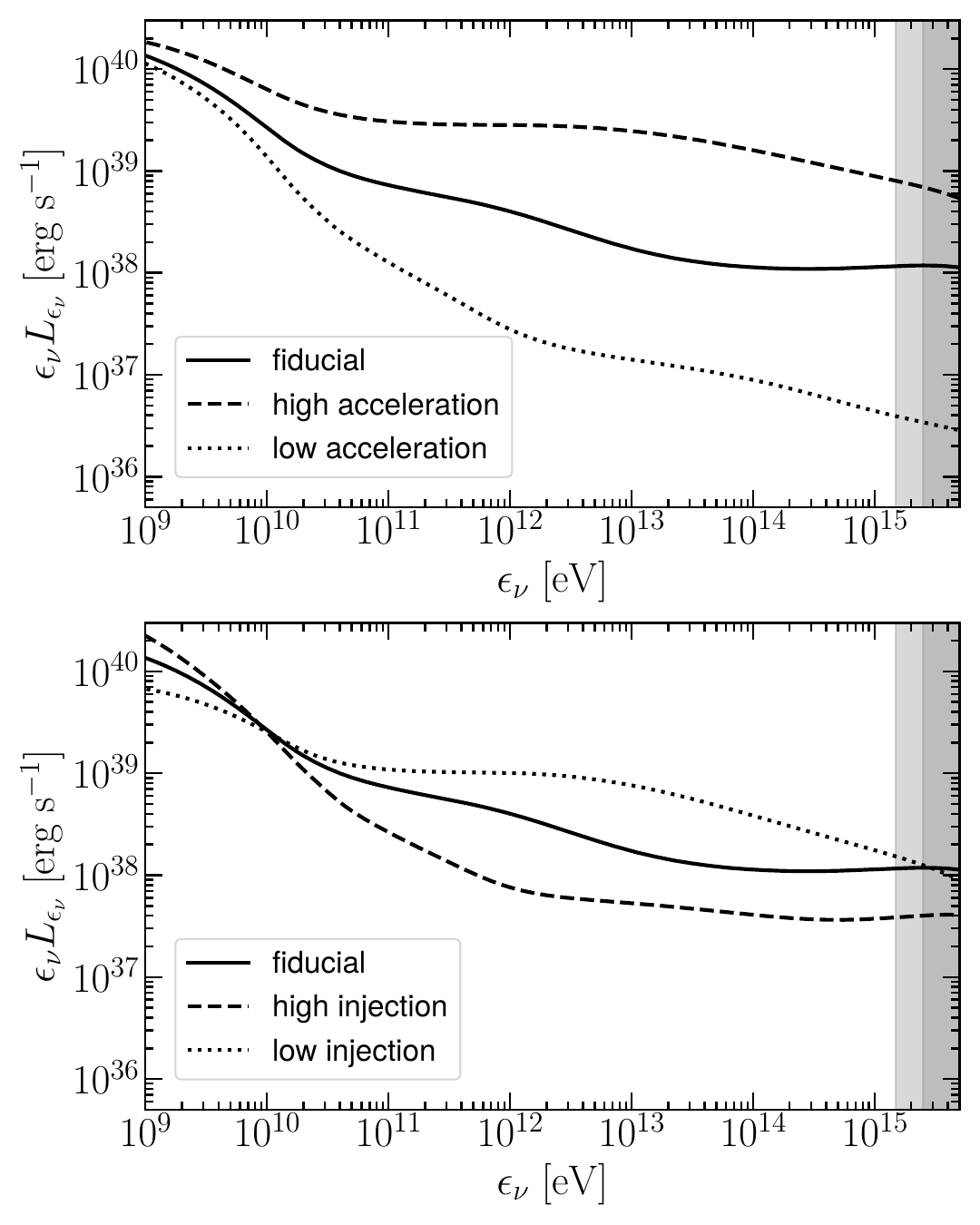}    \caption{The dependence of the high-energy neutrino SEDs on the parameters. Upper: acceleration efficiency dependence with 
    $\eta_{\rm accel} = 10^{-5}$ (low), $3 \times 10^{-5}$ (fiducial), and $10^{-4}$ (high).
    Lower: injection efficiency dependence with 
    $\eta_{\rm inj} = 3 \times 10^{-4}$ (low), $10^{-3}$ (fiducial), and $3 \times 10^{-3}$ (high).
    }
    \label{fig:Neutrino_SED_ave_sum_flux_woObs_Comp}
\end{figure}

Our simple phenomenological model for the CR injection and acceleration includes only two parameters, $\eta_{\rm inj}$ and $\eta_{\rm accel}$. As shown in the previous subsections, our parameter choice produces the neutrino emission with a luminosity of 
$\sim 6 \times 10^{38} {\rm erg} ~{\rm s}^{-1}$ in the energy range $ 1~ {\rm TeV} \le \epsilon_{\nu} \le 100~ {\rm TeV}$.   
As shown in the next section, this luminosity can be consistent with the extragalactic diffuse neutrinos.
Here, we examine the dependence of the neutrino spectrum on those free parameters.
The upper panel of Figure \ref{fig:Neutrino_SED_ave_sum_flux_woObs_Comp} shows that the neutrino SED becomes harder as the acceleration efficiency increases. 
This trend seems a natural consequence. The total neutrino luminosity is enhanced by 
$\sim 8$ and $\sim 0.04$ in $ 1~ {\rm TeV} \le \epsilon_{\nu} \le 100~ {\rm TeV}$ for high and low acceleration efficiency models, respectively.
An important finding is that the neutrino spectra are difficult to express by a single power-law distribution, which is an outcome of the inhomogeneous and time variable structure of the accretion flows and outflows in our computations.

The lower panel of Figure \ref{fig:Neutrino_SED_ave_sum_flux_woObs_Comp} displays the dependence on the injection efficiency. 
Below $\sim 10^{11}$ eV, the spectrum becomes softer with increasing $\eta_{\rm inj}$. Given $U_{\rm th}$, a higher CRP injection implies a lower energy gain per particle.
Therefore, in the high energy ($\gtrsim 10^{10}$ eV) region, a smaller $\eta_{\rm inj}$ increases the neutrino flux.
The total neutrino luminosity is enhanced by 
$\simeq 0.35$ and $\simeq 1.6$ in $ 1~ {\rm TeV} \le \epsilon_{\nu} \le 100~ {\rm TeV}$ for high and low injection efficiency models, respectively.

In summary, the appearance of the relatively soft neutrino SED profile, which is one of the main conclusions of this paper, is not sensitive to the choice of the free parameters in our models.
However, another sub-grid model for the acceleration or injection may change the spectrum significantly.

\section{Discussion} \label{sec:discussion}
\subsection{Diffuse neutrino intensity}
Although our purpose in this paper is to demonstrate the effects of non-steady and non-uniform CRP acceleration and injection rather than the quantitative fitting of the observed neutrino spectra,
we compare the moderately flat neutrino SEDs  obtained by our global CRP acceleration simulations with the extragalactic diffuse neutrino spectrum detected with IceCube \citep{IceCube_diffuse_2020, IceCube_diffuse_2024}.
The diffuse neutrino intensities are calculated by previous studies \citep[e.g.,][]{Alvarez-Muniz_Meszaros_2004, Murase_etal_2014, Kimura_etal_2015, Kimura_etal_2021}:
\begin{eqnarray}
    \epsilon_{\nu}^2 \Phi_{\epsilon_{\nu}} = \frac{c}{4\pi H_0} 
    \int_0^{z_{\rm max}} \frac{\dd z}{\sqrt{\Omega_{\rm m}(1+z)^3 + \Omega_{\Lambda}}} \nonumber \\
    \times \int_{L_{\rm min}}^{L_{\rm max}} \dd L_{{\rm H}_{\alpha}} \dv{n_*}{L_{{\rm H}_{\alpha}}} \left(\frac{\epsilon_{\nu}}{\hat{\epsilon}_{\nu}}\right)^2 \hat{\epsilon}_{\nu} L_{\hat{\epsilon}_{\nu}}, \label{eq:diffuse_neutrno}
\end{eqnarray}
where $z$ is the cosmological redshift, $H_0 = 7 \times 10^6 ~{\rm cm} ~ {\rm s}^{-1} ~ {\rm Mpc}^{-1}$ is the Hubble constant, and $\Omega_{\rm m} \approx 0.3$ and $\Omega_{\Lambda} \approx 0.7$ are the cosmological density parameters for  matter and dark energy, respectively.
The neutrino energy measured in the rest frame of LLAGN is $\hat{\epsilon}_{\nu} = (1+z) \epsilon_{\nu}$.
For the luminosity function, we adopt the ${\rm H}_{\alpha}$ luminosity function of nearby LLAGNs $\dd n_* / \dd L_{{\rm H}_{\alpha}}$ \citep{Ho_2008}, which is fitted and adopted in \cite{Kimura_etal_2015} with assuming no dependence on $z$ as follows:
\begin{eqnarray}
    \dv{n_*}{L_{{\rm H}_{\alpha}}} 
    = \frac{n_* / L_{{\rm H}_{\alpha}}}{(L_{{\rm H}_{\alpha}}/L_*)^{s_1} + (L_{{\rm H}_{\alpha}}/L_*)^{s_2}},
\end{eqnarray}
where $L_{{\rm H}_{\alpha}}$ is the luminosity of  ${\rm H}_{\alpha}$ emission, $L_* = 10^{38} ~{\rm erg} ~ {\rm s}^{-1}$, $n_* = 1.3 \times 10^{-2} ~{\rm Mpc}^{-3}$, $s_1 = 1.64$, and $s_2 = 1$.

The bolometric radiation luminosity can be written as $L_{\rm bol} \sim 80~ L_{{\rm H}_{\alpha}}$.
For simplicity, we assume that the neutrino luminosity generated by $pp$ interactions is proportional to ${\dot M}^2$, which is a reasonable approximation in the RIAF limit.
The radiative efficiency $\eta_{\rm rad} = L_{\rm bol}/({\dot M} c^2)$ is approximated as $\eta_{\rm rad} = \eta_0 ({\dot m}/10^{-2})^{q}$
with ${\dot m} = {\dot M} c^2/L_{\rm Edd}$.
Here, we adopt $\eta_0 = 8\times 10^{-2}$ and $q=1/2$ (i.e., $L_{\rm bol} \propto {\dot m}^{3/2}$), which can approximately reproduce the radiative efficiency computed in, e.g., \cite{Xie_Fuan_2012} and \cite{Yarza_etal_2020}.
Then, the relation between the expected neutrino SEDs in the LLAGN rest frame and the bolometric luminosity can be obtained as follows: 
\begin{eqnarray}
    \hat{\epsilon}_{\nu} L_{\hat{\epsilon}_{\nu}}
    &=& \hat{\epsilon}_{\nu} L_{\hat{\epsilon}_{\nu}} ({\dot m} = 10^{-2}) \left( \frac{\dot m}{10^{-2}}\right)^2 \nonumber \\
    &=&  \hat{\epsilon}_{\nu} L_{\hat{\epsilon}_{\nu}} ({\dot m} = 10^{-2}) \left( \frac{L_{\rm bol}}{10^{-2} \eta_0}\right)^{2/(q+1)}.
\end{eqnarray}
The luminosity is normalized at $\dot{m}=10^{-2}$ with our simulation results. Assuming the same spectral shape, we integrate the contributions from $\dot{m}=10^{-7}$ to $10^{-2}$, which is the same as the definition of LLAGNs in \citet{Kimura_etal_2015}. 

\begin{figure}
	\includegraphics[width=\columnwidth]{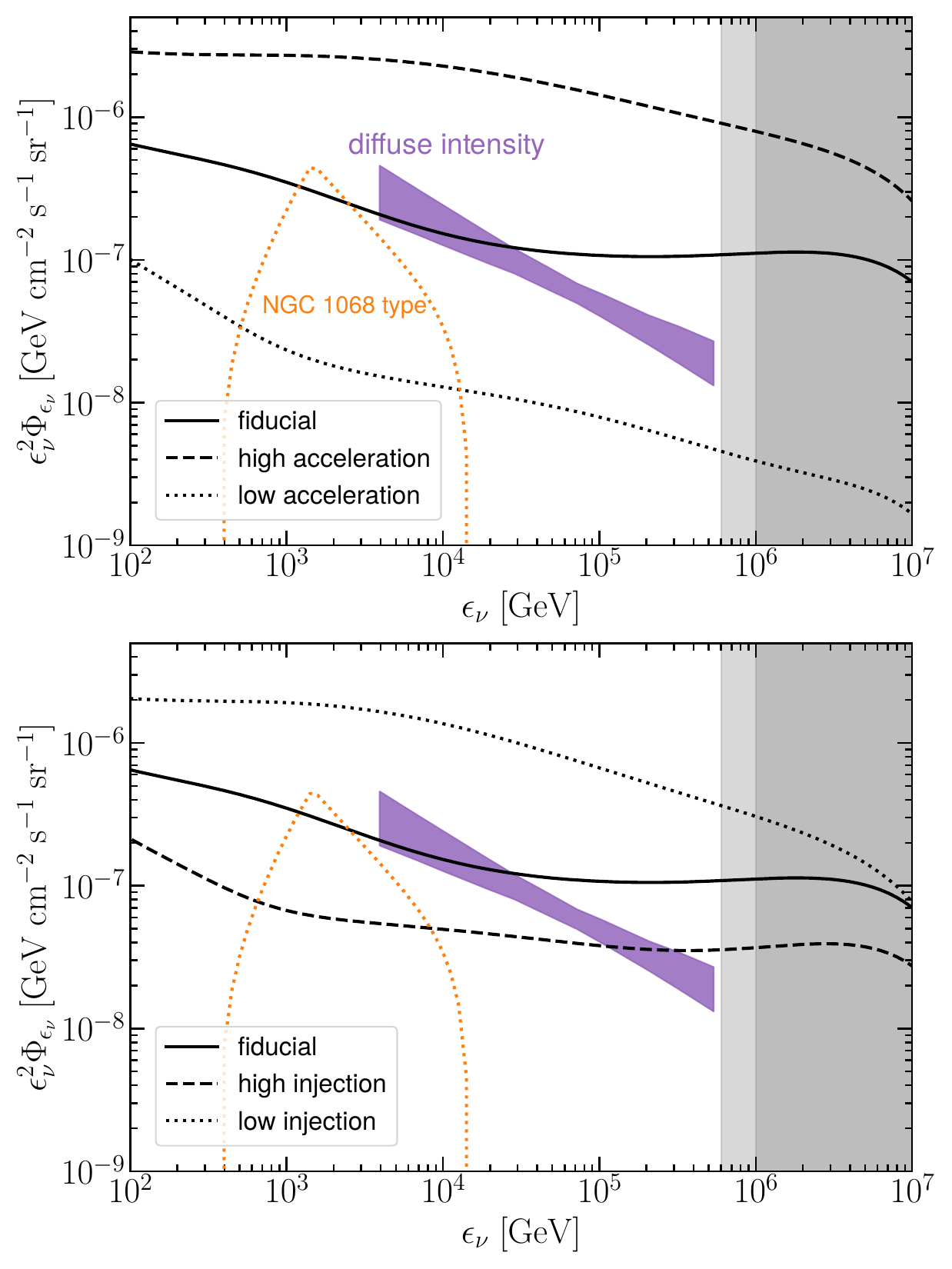}    \caption{The diffuse all-flavor neutrino intensity estimated with our models.
    The dependence of the acceleration efficiency (top) and the injection efficiency (bottom) are shown. The purple-shaded region represents the reconstructed intensity from IceCube observations \citep{IceCube_diffuse_2024}, 
    where the equal flavor ratio is assumed.
    The orange dotted line indicates the expected contribution from NGC 1068-like (i.e., Seyfert-type) AGNs.}
    \label{fig:diffuse_neutrino}
\end{figure}

For reference, we also estimate the diffuse neutrino intensity from the NGC 1068-like sources using the observed neutrino flux of NGC 1068 \citep{IceCube_NGC1068_2022},
assuming a density of $\sim 5 \times 10^{-5}~{\rm Mpc}^{-3}$ of quiescent AGNs at $L_{\rm bol} \sim 10^{45} ~ {\rm erg ~ s}^{-1}$ \citep{Aversa_etal_2015}. 
We adopt $L_{\rm bol}$ estimated by the near-infrared observation of NGC 1068 with GRAVITY \citep{Garvity_etal_2020}.

Figure \ref{fig:diffuse_neutrino} displays the resulting diffuse all-flavor neutrino intensity. 
Since the detailed SED fitting is beyond the scope of this paper and our sub-grid model of the particle acceleration is still primitive, 
we instead discuss the key features of the SED profile by briefly comparing it with the observed diffuse neutrino intensity \citep{IceCube_diffuse_2024}.
One can find that the power-law indices of the resulting SEDs depend on the parameters. 
We do not focus on the spectra above $\sim 5 \times 10^2$ TeV, where our CRP advection approximation and neglect of the $p\gamma$ process are inappropriate.
The fiducial model seems roughly consistent with the observed flux level.
Suppose we decrease the fraction of the AGN population contributing to the neutrino background. In that case, the high-acceleration/low-injection models can be consistent with the observed spectrum owing to their relatively soft spectra.
Another possible scenario is the combination of the NGC 1068-like components and another harder and dimmer component like the high-injection model in Figure \ref{fig:diffuse_neutrino}. Such two-population models have been discussed in previous studies \citep[e.g.][]{Kimura_etal_2015,Kimura_etal_2021}.

\subsection{CR luminosity}
It will be worth mentioning that our fiducial model indicates the outflow CRP luminosity of $L_{\rm CRP} \sim 10^{43}  ~ {\rm erg} ~{\rm s}^{-1} \left(\frac{M_{\rm BH}}{10^8 M_{\odot}}\right) \left(\frac{{\dot M}}{10^{-2}L_{\rm Edd}/c^2}\right)^2 ~ {\rm erg}$. 
For example, in Sgr A$^*$, $M_{\rm BH} \simeq 4 \times 10^{6} M_{\odot}$ and ${\dot M} \sim 10^{-7} - 10^{-5} L_{\rm Edd}/c^2$ \citep[e.g., ][]{EHTC_2022a_SgrA, EHTC_2022e_SgrA}, we obtain $L_{\rm CRP} \sim 10^{31-35} ~ {\rm erg} ~{\rm s}^{-1}$. 
This expected CRP luminosity is much less than the estimated CR luminosity in Sgr A$^*$ \citep[$\sim 10^{41} ~ {\rm erg} ~{\rm s}^{-1}$,][]{Dogiel_etal_2002}.
Even with a more optimistic parameter choice, the accretion flow model for the CRP acceleration seems unlikely for the main CR source in Sgr A$^*$, except the temporal activity model with a high mass accretion onto Sgr A$^*$ in the past with a duration of $10^{6-7}$ yrs \citep{HESS_2016Nature, Fujita_etal_2017}.\footnote{It should be noted that this scenario is under debated. The Fermi bubble \citep{Su_etal_2010_FermiBubble} and the eROSITA bubble \citep{Predehl_etal_2020Nature_eROSITAbubble} can be evidence of the past activity of Sgr A$^*$, but they can also be explained by persistent emissions due to the CR-driven Galactic fountain flows \citep{Shimoda_Asano_2024}.}

On the other hand, in M82, the supernova-origin CR luminosity estimated from the star formation rate  ($\sim 10~M_\odot \mbox{yr}^{-1}$) is $\sim 3\times 10^{41}  ~ {\rm erg} ~{\rm s}^{-1}$ \citep{Heinz_Zweibel_2022}. 
While the mass of the central supermassive black hole in M82 is estimated to be $\sim 3 \times 10^{7} M_{\odot}$ \citep{Gaffney_etal_1993_M82mass}, the mass accretion rate is still unknown. 
If we assume that ${\dot M} \sim 10^{-2}  L_{\rm Edd}/c^2$, the CR luminosity in the outflow from the AGN disk can dominate that from the supernova origin. 
If the RIAF model for the neutrino background is valid, CRs from the central engine of AGNs may have a significant impact on the evolution of their host galaxies. 

\section{Summary}  \label{sec:summary}

We have carried out the computation of the CRP acceleration and high-energy neutrino emission taking into account the inhomogeneous and time-dependent structure of accretion flows and outflows around a supermassive black hole with a global GRMHD simulation. 
We assume that CRPs are advected with the background fluid and accelerated by the kinetic-scale turbulence and generate neutrinos via $pp$ interaction.
The CRP acceleration and injection processes in the sub-grid scale in our MHD simulation are calculated with a simple phenomenological model with two parameters.
The neutrino SEDs exhibit softer profiles compared to those of single-zone models.
This arises from the non-steady acceleration and injection efficiencies, which evolve with the background fluid. We have obtained a variety of the CRP spectrum. The superposition of contributions from various CRP SEDs results in a flat spectrum in the wide energy range.
The neutrino SEDs are predominated by contributions of outflow CRPs, which undergo acceleration and subsequently collide with thermal protons in the accretion flow during the early phase of their trajectories before escaping from the simulation domain.
The neutrino SED becomes harder as the acceleration efficiency increases.
On the other hand, when the injection efficiency is higher, the neutrino SEDs are dimmer and softer in mildly high-energy range ($\gtrsim 10^{11} ~ {\rm eV}$).
Our result shows that CRs advected with the outflow can significantly contribute to CRs in the host galaxy.

Given an AGN population as neutrino sources,
within the framework of our phenomenological model, we have found a typical parameter set for the CRP acceleration and injection processes to 
be roughly consistent with the observed flux level of the extragalactic neutrino background.
Alternatively, the diffuse neutrino spectrum may be explained by the superposition of the contributions from NGC 1068-like AGNs and LLAGNs, given our high-injection model.
A more quantitative comparison including the effects of $p \gamma$ interactions, spatial diffusion of CRPs, more sophisticaed model of CRP acceleration, strongly magnetized accretion flow model (i.e., MAD), and full relativity are left as future work.

\begin{acknowledgments}
We acknowledge H.R. Takahashi and K. Ohsuga for helpful discussion on the GRMHD simulations using \texttt{UWABAMI} code. 
We also thank S. S. Kimura, K. Murase and J. Shimoda for the fruitful discussion. 
The numerical simulations were carried out on the XC50 at the Center for  Computational Astrophysics, National Astronomical Observatory of Japan. 
 This work was supported by the joint research program of
the Institute for Cosmic Ray Research (ICRR), the University
of Tokyo, and JSPS KAKENHI grant Nos. JP22H00128,  
JP23K03448, JP23H00117, JP24K00672 (T.K.), JP22K03684, JP23H04899, and JP24H00025 (K.A.).
This work was also supported by MEXT as 
 “Program for Promoting Researches on the Supercomputer Fugaku” (Structure and Evolution of the Universe Unraveled by Fusion of Simulation and AI; Grant Number JPMXP1020230406; Project ID hp240219), by the HPCI System Research Project (Project ID: hp240054).
\end{acknowledgments}

%






\newpage
\appendix
\section{Timescales of hadronuclear and photohadronic interactions} \label{sec:app1}

In this paper, the effects of photohadronic interactions are not included.
To assess their influence on the resulting neutrino SEDs, we roughly estimate the energy range of neutrinos 
for
which $p\gamma$ interaction is non-negligible.
For simplicity, we adopt a single-zone approximation.

The inverse of the timescales for $pp$ and $p \gamma$ interactions are described as follows \citep[e.g.,][]{Kimura_etal_2015}:
\begin{eqnarray}
    t_{pp}^{-1} &=& \tilde{n}_{\rm MHD} \sigma_{pp} (\gamma_{\rm CRP}) c K_{pp}, \label{eq:t_pp_inverse} \\
    t_{p\gamma}^{-1} &=& \frac{c}{2\gamma_{\rm CRP}^2} E_{\rm rel} \Delta E_{\rm rel} \sigma_{p \gamma} (E_{\rm rel}) K_{p \gamma} (E_{\rm rel}) \int_{E_{\rm rel}/(2\gamma_{\rm CRP}) }^{\infty} \dd E_{\gamma} \frac{n_{\gamma} (E_{\gamma})}{E_{\gamma}^2},
    \label{eq:t_pg_inverse}
\end{eqnarray}
where $\gamma_{\rm CRP}$ is the Lorentz factor of CRPs, $E_{\gamma}$ is the photon energy in the laboratory frame 
, $N_{\gamma} (E_{\gamma})$  is the isotropic photon number density in the laboratory frame, $E_{\rm rel}$ is the photon energy in the CRP rest frame, $\tilde{n}_{\rm MHD} (\sim 10^{11} ~{\rm cm}^{-3})$ is the typical value of the thermal proton density inside the accretion flow in our model, $\sigma_{pp}$ and $\sigma_{p \gamma}$ are the cross section for $pp$  and $p \gamma$ interaction, $K_{pp} (\simeq 0.5)$ and  $K_{p \gamma} (\simeq 0.2)$ are the inelasticity of the $pp$ and $p \gamma$ interactions, respectively. 
For the $p \gamma$ interactions, rectangular approximation \citep{Stecker_1968} is assumed and $E_{\rm rel} \sim 3 \times 10^8 {\rm eV}$, $\Delta E_{\rm rel} \sim 2 \times 10^8 {\rm eV}$, and $\sigma_{p \gamma} \sim 5 \times 10^{-28} {\rm cm^2}$.

We assume a simplified form of the cross section $\sigma_{pp} \simeq (34.3  + 1.88 L + 0.25 L^2) [1 - (E_{pp, {\rm thr}}/\gamma_{\rm CRP} m_{\rm p} c^2)^4] \times 10^{-27} ~ {\rm cm}^{2}$ for $\gamma_{\rm CRP} m_{\rm p} c^2 \ge E_{pp, {\rm thr}}$, where $L = \log(\gamma_{\rm CRP} m_p c^2/ 10^3 ~{\rm GeV})$ and $E_{pp, {\rm thr}} = 1.22 ~{\rm GeV}$ \citep{Kelner_2006}. 
We can, therefore, roughly estimate $t_{pp}^{-1} \sim 10^{-6} (34.3  + 1.88 L + 0.25 L^2) [1 - (E_{pp, {\rm thr}}/\gamma_{\rm CRP} m_p c^2)^4] ~{\rm s}^{-1}$.

With respect to $p \gamma$ interactions, for the 
integral in Equation (\ref{eq:t_pg_inverse}), we assume that the power-law distribution of the $n_{\gamma} = n_{-3} (E_{\gamma}/10^{-3} {\rm eV})^{s}$ with $s \sim -2$ and $4 \pi (10r_{\rm g})^2 c E_{\gamma}^2 n_{-3}(E_{\gamma})|_{E_{\gamma} = 10^{-3} {\rm eV}} \sim 10^{40} ~{\rm erg} ~{\rm s}^{-1}$, which roughly agrees with the photon SED of the model C (i.e., $M_{\rm BH} = 10^8 M_{\odot}$ and ${\dot M} = 10^{-2} L_{\rm Edd}/c^2$) in \cite{Kimura_etal_2015}.
The last integral 
is described as 
$\simeq 2 \times 10^{-37} \gamma_{\rm CRP}^3 ~ {\rm cm}^{-3}~ {\rm eV}^{-1}$. 
We, then, obtain $t_{p \gamma}^{-1} \sim 2 \times 10^{-11} \gamma_{\rm CRP}~\mbox{s}^{-1}$.

As a consequence, $p \gamma$ interactions can dominates, i.e., $t_{p \gamma}^{-1} \gtrsim t_{pp}^{-1}$, for $\gamma_{\rm CRP} m_p c^2 \gtrsim 5 \times 10^{16} ~ {\rm eV}$.
As the neutrino energy is $\sim 0.05 \times \gamma_{\rm CRP}m_{\rm p} c^2$,
our neutrino SEDs will be uncertain in the energy range $\gtrsim 2 \times 10^{15} ~ {\rm eV}$, i.e., the shaded region in Figure \ref{fig:Neutrino_SED_ave_sum_flux_woObs}.

\bibliography{GRRT,GRRT2,RHD,RIAF,neutrino}
\bibliographystyle{aasjournal}

\end{document}